\documentclass[runningheads]{llncs}
\usepackage[T1]{fontenc}
\usepackage{graphicx}
\usepackage{amsmath}
\usepackage{color}
\usepackage[hidelinks]{hyperref}

\newcommand\blfootnote[1]{%
  \begingroup
  \renewcommand\thefootnote{}\footnote{#1}%
  \addtocounter{footnote}{-1}%
  \endgroup
}
\begin{document}
\title{Simulating Validity: Modal Decoupling in MLLM Generated Feedback on Science Drawings}
\titlerunning{Modal Decoupling in MLLM Feedback on Science Drawings}
\author{Arne Bewersdorff\inst{1} \and
Nejla Yuruk\inst{2} \and
Xiaoming Zhai\inst{1}}
\authorrunning{A. Bewersdorff et al.}
\institute{University of Georgia, AI4STEM Education Center, Athens GA 30602, USA\\
\email{xiaoming.zhai@uga.edu} \and
Gazi University, Department of Mathematics and Science Education, Ankara, 06560, Turkiye}
\maketitle
\blfootnote{\itshape Accepted as AIED Short Paper 2026, Seoul, South Korea. Submission \#1147.}
\begin{abstract}
In science education, students frequently construct hand-drawn visual models of scientific
phenomena. 
These drawings rely on a visual structure where information is encoded through visual objects, 
their attributes, and relationships. Multimodal large language models (MLLMs) are increasingly 
used to generate feedback on students' hand-drawn scientific models. However, the validity 
of such feedback depends on whether model claims are grounded in the specific visual evidence of 
the student drawing. This study uncovers grounding failures, consistent with modal decoupling, in 
off-the-shelf MLLM feedback, where outputs remain pedagogically plausible in form while
contradicting 
the drawing or treating depicted elements as missing. Using $N = 150$ middle school drawings from a 
kinetic molecular theory unit spanning five modeling tasks and three competence levels, we generated
$N = 300$ feedback 
instances with GPT-5.1. All outputs were coded for four grounding error types: object mismatch,
attribute mismatch, 
relation mismatch, and false absence. Grounding failures were common: 41.3\% of feedback instances
contained at 
least one error. An inventory-list-first workflow reduced several error categories and lowered the
overall error 
rate, but it did not resolve the underlying limitation: approximately one in three outputs remained
flawed, with 
false absence as the dominant failure mode. Moreover, feedback that appears visually grounded
offered little diagnostic 
value for identifying invalid instances. The findings indicate that modal decoupling is a
substantial 
limitation and that valid feedback will require grounding mechanisms beyond common prompting
strategies.

\end{abstract}

\section{Introduction}

In science education, students frequently construct hand-drawn visual models of scientific
phenomena. For 
feedback generated by Multimodal Large Language Models (MLLM) to be valid, the MLLM must accurately
process 
not only the visuals (like molecules and arrows) but also their attributes (like color, shape) and
how objects 
relate to each other in space. If an MLLM correctly identifies a drawing showing ``water particles''
but fails to map 
the student's added arrow indicating kinetic energy, any subsequent feedback regarding the student's
drawing is not grounded 
in the visual evidence of the student's drawing and cannot validly capture the student's scientific
thinking.  
Feedback grounded in the students' drawing is often hindered by modal decoupling, a failure of
visual grounding where 
generated feedback does not reliably bind specific visual elements, their attributes, and their
relations to what is 
actually present in the student's drawing. Prior work suggests that modal decoupling in MLLMs may be
influenced by architectural 
and training dynamics, including geometric separation between modalities in latent space, unimodal
optimization bias toward text, 
and weakening visual constraint during long-form generation \cite{ref_liang}.

MLLMs balance parametric memory (internal training data) against visual evidence (the specific
input) during generation. When visual signals are complex or contradictory, the model minimizes
perplexity by defaulting to its strong linguistic priors (e.g., generic feedback). In this paper, we
demonstrate that modal decoupling, which has been linked to architectural and optimization
constraints in prior work, can lead to invalid feedback when using off-the-shelf pre-trained MLLM
like GPT-5.1 and cannot be reliably mitigated by a widely used inventory-list-first prompting
workflow. By analyzing $N = 150$ student drawings, we show that directly generated feedback
frequently exhibits grounding errors consistent with modal decoupling and produces flawed feedback.
Generating an inventory-list-first, a strategy proposed in prior research \cite{ref_yan}, can
increase the extent to which feedback references the student's drawing, but it does not suppress
modal decoupling. As a result, grounding errors persist. Overall, neither direct image-based
feedback nor inventory-list-first feedback yields consistently valid grounded feedback; both
workflows frequently produce feedback that is formally plausible but not valid with respect to the
visual evidence in the student's drawing. Our contribution is an empirical characterization and
comparison of grounding failures under two deployment-realistic prompting workflows; mechanistic
accounts are discussed as plausible contributors rather than causally identified drivers.

\section{Theoretical Background}

\subsection{Multimodal Representational Competence \& Feedback in Science Education}

Scientific phenomena often involve abstract or microscopic processes, such as particle motion,
energy transfer, or chemical reactions, that are challenging to convey purely through text.
Therefore, science education relies heavily on multimodal construction, where students demonstrate
understanding by seamlessly integrating visual models (e.g., drawings) with explanatory meaning
through entities, their attributes, and their relations. Achieving this integration depends on
representational competence, which is not merely the inclusion of visual elements, but the ability
to establish a successful semiotic connection between what is drawn and what it is intended to
represent (e.g., particles, motion, mechanisms) \cite{ref_ainsworth}.

Students' drawings can serve as a basis for assessing their ideas and competencies, providing  
a valuable source of evidence for revealing their thinking. based on the assessment, providing feedback on students' 
drawings can directly support their scientific understanding. 
However, for this feedback to be effective, it must be grounded in the student's drawing.Feedback on such work is 
valid only if its claims are grounded in the
student's drawing. Grounding invalidity can occur in two distinct ways: (a)
contradiction, where feedback reports entities, attributes, or relations that the drawing does not 
support, and (b) false absence, where feedback explicitly treats a depicted element as missing or 
requests its addition. Valid feedback therefore requires not only detecting visual elements, but
binding feedback claims to the actual student's depicted objects, attributes, and relations.

\subsection{MLLM Processing: Geometric and Optimization Constraints in Modal Decoupling}

The challenge of generating valid feedback in multimodal systems is rooted in the architectural
tension between visual perception and textual reasoning. While 
MLLMs are designed to integrate different types of data, the current ``coupled'' paradigm, where
visual and textual features are processed into a shared latent space, encounters fundamental
hurdles. In pre-trained models, these vector-geometric and optimization constraints are built into
the architecture, potentially leading to misalignment between the model's visual inputs and its
generated textual outputs. These constraints have been associated with modal
decoupling \cite{ref_bai}: phenomena where the generated response fails to align with the visual
content. We operationalize modal decoupling through a structured error coding of grounding failures
in model testual outputs: when feedback contradicts the drawing (E1--E3) or treats depicted
elements as missing (E4). Potential plausible contributors, rather than directly verified causal processes, to modal decoupling 
include geometric separation, unimodal optimization bias, and grounding decay.

\textit{Geometric Separation.} 
Embeddings from different modalities (like text and images), when L2-normalized, can concentrate 
in narrow regions (`cones') of the representation space \cite{ref_liang}. This geometry is often 
discussed in relation to the modality gap, where image and text representations form separated 
clusters. Yi et al.\ \cite{ref_yi} argue that a persistent gap arises in settings where representations 
are effectively constrained to distinct subspaces due to dimension collapse.

This separation between modalities can limit fine grained cross modal alignment: under subspace
separation associated with dimension collapse, paired samples cannot be perfectly aligned, implying
a nonzero mismatch between visual and textual representations. This weak cross-modal alignment is expected to surface as
feedback that either fills in missing visual specifics with plausible defaults or fails to reflect
fine-grained attributes and relations present in the image.

\textit{Unimodal Bias.} 
Beyond geometric separation in the vector space, 
MLLMs face challenges during the optimization process known as Unimodal Bias \cite{ref_zheng}. This bias 
towards one modality occurs when a network relies more on the modality that allows it to reduce error most 
quickly during the training (gradient descent). Typically, this is the textual modality, which is more 
structured and symbolic than high-dimensional visual data.

If the model exploits a successful pattern in the text alone, the gradients for the visual branch
diminish (Gradient Starvation) \cite{ref_zheng}. As a result, the vision encoder remains
under-optimized.
The textual modality dominates the decision boundary, effectively suppressing the signals from the
``weak'' visual modality \cite{ref_bai}. This can result in attribute and relation
hallucinations. In feedback generation, this bias could manifest as over-reliance on generic,
high-probability pedagogical statements that remain plausible even when they do not match the
specific details of the student drawing.

\textit{Visual Grounding Decay.} 
As the model predicts one word after another, the attention mechanism's connection 
to the original visual feature vectors weakens over time (Visual Grounding Decay) \cite{ref_chung}. In a 
transformer architecture, each newly generated word consumes a portion of the model's attention. Over a long text sequence, 
the influence of the static visual embeddings is often diluted by the growing textual context \cite{ref_favero}. 

Once a model predicts a specific label or conclusion based on its internal textual patterns, the
autoregressive nature of the system forces it to remain consistent with that choice. Any following
analysis of the input is generated to match that initial judgment. This mechanism effectively
decouples the visual evidence from the final explanation, resulting in a system that ``fills in''
(hallucinates) details based on general textual patterns. The reasoning module detaches from the perceptual
data to maintain linguistic consistency, leading to outputs that appear correct but are not grounded
in specific visual evidence. A practical signature is that early statements appear image-tied, but
later parts of the feedback drift toward self-consistency and generic reasoning rather than
continued reference to the visual evidence. For
instance, if a student labels a simple circle as an ``animal cell'' but omits the nucleus, the
separation of the model's textual subspace from the visual subspace can trigger the high-probability
prior that animal cells contain a nucleus. Despite the visual evidence of an empty circle, the model
praises the ``depiction of the nucleus.'' The model detaches from the visual data to maintain
linguistic consistency, resulting in a simulation of valid feedback rather than a grounded
assessment.

In summary, the decoupling of multimodal features has been linked to at least three distinct
technical constraints: (1) geometirc subspace separation via dimension collapse, (2) unimodal optimization
bias through prior dominance, and (3) autoregressive attention dilution during inference. These
mechanisms of modal decoupling can yield to systemic hallucinations that potentially invalidate the
model's output, demanding strategies to re-align textual reasoning with visual evidence. 

\textit{Inventory-List-First as a common Visual Grounding Mechanism.} To mitigate systemic
hallucinations, research explores interventions across the MLLM lifecycle. Most strategies are
model-centric, focusing on data (diversifying instruction sets), architecture (stronger alignment
interfaces), or training (RLHF) \cite{ref_bai}. However, in science education, where practitioners
often rely on fixed, pre-trained MLLMs like GPT-4v or GPT-5.1, these methods are often infeasible
\cite{ref_bewersdorff,ref_yin}. Consequently, inference-centric strategies that modify the reasoning chronology
of of-the-shelf-MLLM are more applicable. Among these, the inventory-list-first workflow emerged as a robust
mechanism to increase modal alignment during the generation process.
Inventory-list-first is a serialized reasoning strategy, structurally analogous to Chain-of-Thought
(CoT) \cite{ref_yan} prompting. In this paradigm, the model is compelled to resolve and verbalize
specific visual tokens, such as student-drawn objects, attributes or relationships, before
initiating pedagogical evaluation. Theoretically, this serialization may increase the likelihood to
successfully address the technical constraints of modal decoupling by bridging the modality gap. By
verbalizing visual features first, the model maps abstract visual embeddings to discrete textual
tokens, aiming to migrate visual evidence into the textual subspace to overcome dimension collapse.
This chronology is aimed at suppressing Unimodal Bias and Prior Dominance by forcing an explicit
observation phase that prevents the model from defaulting to statistical priors without engaging the
visual input. Furthermore, the resulting inventory can serve as an anchor in the context window,
mitigating Attention Dilution and preventing Post-Hoc Rationalization by ensuring evidence remains
mathematically accessible during the evaluation phase. These mechanisms motivate why inference
strategies like inventory lists that alter reasoning chronology might reduce observable modal
decoupling, but they do not imply that any single prompt will eliminate them; our study evaluates
this empirically for two commonly used workflows.

\subsection{Construct Validity and the Risk of Fabricated Utility}

The technical decoupling of multimodal features poses a fundamental threat to validity of student
assessment. When feedback outcomes are influenced by a model's training priors rather than the
student's specific drawn input, the system introduces construct-irrelevant variance. For example, if
an MLLM tells a student to add an element that is already present, it reflects a failure to
incorporate depicted evidence into the feedback claim. This results in feedback which still appears
pedagogically sound in form but is invalid in substance. This disconnect represents a fundamental
threat to validity: the feedback loop is closed not by the student's evidence, but by the model's
own probabilistic expectations.

If modal decoupling produces feedback that is formally plausible but not grounded in the student's
drawing, a follow-up question arises: can the validity of feedback be detected from its surface
properties? Automated quality filters might assume that feedback containing more concrete visual
references (e.g., physical-term density), more spatial and relational language, or epistemic hedging
is more likely to be grounded in the actual drawing.

\section{Research Questions}

The present study is guided by the following research questions, which examine grounding failures as
signs of modal decoupling in MLLM-generated feedback on students' scientific drawings.

\noindent\textbf{RQ1.} To what extent does off-the-shelf MLLM-generated feedback on student scientific drawings exhibit grounding failures?

\noindent\textbf{RQ2.} Which forms of grounding failures are most prevalent, and how do they vary across modeling tasks and competence level?

\noindent\textbf{RQ3.} To what extent do grounding failures persist under an inventory-list-first workflow intended to mitigate grounding failures during feedback generation?

\noindent\textbf{RQ4.} Do automated linguistic indicators of visual grounding (e.g., word count, physical noun density, hedging) predict lower grounding failures (i.e., fewer contradiction errors and false absence errors), or does decoupling persist despite stronger grounding-oriented linguistic profiles?

\section{Method}

\subsection{Sample}

The dataset for this study is based on \cite{ref_zhai2022} and consists of $N = 150$ student-generated scientific drawings created by
middle school students in a curriculum unit on the Kinetic Molecular Theory (KMT). Students were
asked to draw visual models of unobservable particle level mechanisms to explain macroscopic
phenomena, such as gas expansion, condensation, and boiling. The drawings cover five modeling tasks
that differ in scientific abstraction and relational complexity, providing a diverse testbed for
evaluating MLLM grounding (Table~\ref{tab:distribution}): (R1) Red dye diffusion, (R2) balloon
expansion and contraction, (R3) heating a solid, (R4) water condensation, and (R5) boiling water. To
examine how model performance varies with visual complexity, each drawing was pre-classified into
one of three competence levels.

\begin{table}
\centering
\caption{Distribution of student drawings over competency levels and tasks.}\label{tab:distribution}
\begin{tabular}{@{}l@{\hspace{1.2em}}l@{\hspace{1.2em}}l@{\hspace{1.2em}}l@{\hspace{1.2em}}l@{\hspace{1.2em}}l@{\hspace{1.2em}}l@{}}
\hline
Level & R1 & R2 & R3 & R4 & R5 & Total \\
\hline
Level 1 & 7 & 10 & 10 & 6 & 18 & 51 \\
Level 2 & 8 & 6 & 9 & 9 & 10 & 42 \\
Level 3 & 9 & 9 & 10 & 9 & 20 & 57 \\
Total & 24 & 25 & 29 & 24 & 48 & 150 \\
\hline
\end{tabular}
\end{table}

\subsection{Experimental Design}

Based on the dataset of student drawings, we evaluated two commonly used operational workflows for
generating textual feedback on student drawings with the OpenAI Chat Completions API (gpt-5.1;
temperature $= 0$; single-image input via base64; January 16, 2026). The C1 condition prompts the
model to generate feedback immediately. In the C2 condition, the MLLM is prompted to first execute a
visual inventory list of: 1) objects present, 2) attributes of the objects present, and 3) the
relationships between the objects present. This serves as the observational basis for the subsequent
feedback. By systematically comparing the two types of feedback, we analyzed if and how the timing
of reasoning impacts the system's ability to resist modal decoupling, ensure grounding validity, and
maintain robustness against fabrication.

\subsection{Analysis}

To assess the feedbacks' grounding validity, all MLLM-generated feedback for both workflows (C1 and
C2) was coded for modal decoupling. Modal decoupling was operationalized as the presence of one or
more grounding failures, interpreted as evidence failures relative to students' representational
choices (entities, attributes, relations), defined as below. We distinguished two instructionally
meaningful classes: (a) ``false evidence'' errors, where feedback contradicts student's drawing and
thus misdiagnoses what the student represented (E1--E3), and (b) false absence errors (E4), where
feedback explicitly states that an element is missing or instructs the student to add an element
that is already present.

\begin{itemize}
\item \textbf{E1} (Object mismatch: False evidence about represented entities): ``Feedback references an object that is not present in the drawing (false positive) when the object identity is unambiguous.''
\item \textbf{E2} (Attribute mismatch: False evidence about how entities are specified): ``Feedback assigns an attribute (e.g., color, size, state) that contradicts the drawing.''
\item \textbf{E3} (Relation mismatch: False evidence about expressed structure and mechanisms): ``Feedback asserts a spatial or relational structure (e.g., directionality, containment, proximity, causal arrow meaning) that contradicts the drawing.''
\item \textbf{E4} (False absence: False evidence about missing elements): ``Feedback explicitly states that an element is missing or instructs the student to add an element when that element is already present in the drawing.''
\end{itemize}

Each error type was coded as binary ($0 = \text{absent}$, $1 = \text{present}$). Multiple error
types could co-occur within a single feedback. A feedback was classified as flawed if at least one
error type (E1--E4) was present. Two researchers independently coded a subset of 15 drawings (three
per modeling task, 10\% of all drawings). Interrater reliability was high (Cohen's $\kappa = .83$;
94\% agreement).

\begin{figure}
\fbox{\includegraphics[width=0.44\textwidth]{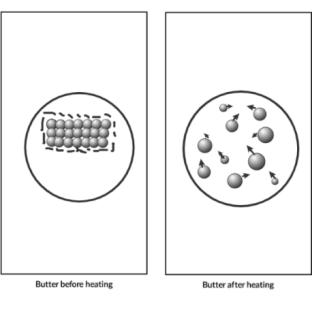}}\hspace{0.05\textwidth}\fbox{\includegraphics[width=0.44\textwidth]{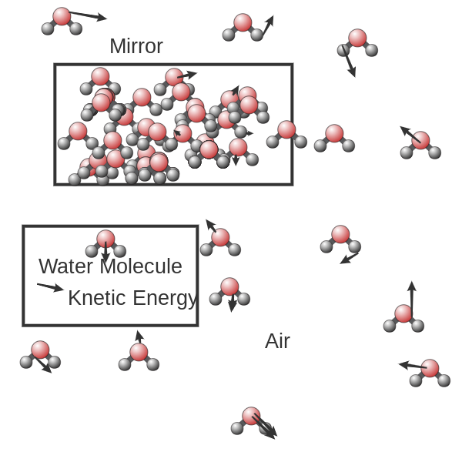}}
\caption{False absence errors (E4) persisting across both workflows. Panel~A (left side, R3, Level~3); Panel~B (right side, R4, Level~3).}\label{fig:examples}
\end{figure}

To illustrate these error types, Figure~\ref{fig:examples} presents two examples. In Panel~A (R3:
heating a solid), a student drew butter particles with motion arrows before and after heating, yet
the feedback requests ``small vibration arrows'', a false absence error (E4), as motion arrows are
already depicted. In Panel~B (R4: water condensation), the student drew water molecules with
differentiated kinetic energy arrows on both the mirror and in the air, yet the feedback claims this
indication is ``missing'' (E4). In both cases, the error persists under the inventory-list-first
workflow (C2), illustrating the ceiling of prompting-based mitigation across different science tasks
and competence levels.

To address RQ1, we estimated the prevalence of flawed feedback separately for each workflow and
pooled across both workflows. Proportions were reported with 95\% confidence intervals computed
using the Wilson score method \cite{ref_agresti}.

To address RQ2, we examined the profiles and boundary conditions of modal decoupling by
distinguishing contradiction-type failures (E1--E3) from false absence failures (E4). Error rates
were computed as the mean number of coded errors per feedback (errors per instance). These rates
were examined as a function of modeling task and student's drawing competence level. Pooled
estimates aggregated feedback across both workflows, and descriptive comparisons were also examined
within each workflow to verify consistency of patterns.

To address RQ3, we tested whether modal decoupling persists under the inventory-list-first workflow.
Because each drawing produced paired feedback (C1 and C2), all comparisons between workflows were
conducted using paired analyses. At the instance level, we compared the proportion of flawed
feedback between workflows. At the error-type level, we compared the prevalence of each error
category using McNemar's test on paired binary indicators (C1 vs.\ C2, $N = 150$ drawings). Effects
were quantified as risk differences expressed in percentage points ($\Delta$~Risk, \textit{pp}),
reflecting the net change in the proportion of instances exhibiting each error type when moving from
C1 to C2. Significance was assessed using McNemar's test \cite{ref_agresti}.

As a descriptive robustness check, we additionally computed error density per generated word (total
errors divided by word count) to examine whether reductions under C2 could be attributed solely to
differences in response length. This measure was used descriptively and not treated as a primary
indicator of feedback grounding validity.

To address RQ4, we examined whether modal decoupling is detectable from feedback text properties. An
automated linguistic analysis was conducted for all feedback, extracting (a) word count, (b)
physical-term density as a proxy for references to concrete visual elements, (c) spatial-term
density to capture relational language, and (d) hedging density as an indicator of epistemic
uncertainty. Physical-term, spatial-term, and hedging densities were computed via lexicon-based
matching and normalized as proportions of the feedback.

Descriptive comparisons were conducted between non-flawed and flawed feedback within each workflow.
We fit cluster-robust logistic regression models \cite{ref_agresti} predicting (a) any flaw
(E1--E4), (b) contradiction-type errors (E1--E3), and (c) false absence errors (E4), controlling for
modeling task, student competence level and workflow condition. Clustering was applied at the
drawing level to account for paired feedback generated for the same drawing.

Finally, to evaluate the practical diagnostic value of linguistic cues, we compared model
discrimination with and without linguistic features using the area under the Receiver Operating
Characteristic curve (Area Under the Curve, AUC, Hosmer et al., 2013) \cite{ref_hosmer}. A baseline
model included task, competence level, and workflow condition; a full model additionally included
linguistic features. The difference in AUC ($\Delta$AUC) was used to assess whether linguistic
properties meaningfully improved the identification of flawed feedback.

\section{Results}

Across $N = 300$ feedback instances, grounding failures were common: 41.3\% of all outputs contained
at least one error. The inventory-list-first workflow reduced several error categories but did not
yield valid feedback; approximately one in three outputs remained pedagogically invalid, with false
absence as the dominant failure mode.

\subsection{Prevalence of Modal Decoupling in Generated Feedback (RQ1)}

Feedback was classified as flawed if it contained at least one coded grounding failure (E1--E4). In
the direct workflow (C1), 49.3\% of instances were flawed (74/150; 95\% CI [41.4\%, 57.3\%]). In the
inventory-list-first workflow (C2), 33.3\% were flawed (50/150; 95\% CI [26.3\%, 41.2\%]). Pooled
across workflows, 41.3\% of feedback were flawed (124/300; 95\% CI [35.9\%, 47.0\%]), indicating
that modal decoupling is common and remains substantial across typical operational workflows.

\subsection{Profiles and Boundary Conditions of Modal Decoupling (RQ2)}

To address RQ2, we distinguished contradiction-type decoupling, where feedback conflicts with the
drawing (E1--E3), from false absence decoupling, where feedback explicitly treats present elements
as missing or requests their addition (E4). Error rates are reported as the mean number of coded
errors per feedback (errors/feedback), with pooled estimates aggregating both workflows.

Across all feedbacks, E4 was the dominant failure mode. In the pooled analyses, the overall E4 rate
was 0.327 errors/feedback, exceeding the pooled contradiction-type rate (E1--E3) of 0.207
errors/feedback. This indicates that the most prevalent form of modal decoupling is false absence,
where feedback explicitly treats already-depicted elements as missing or requests their addition,
rather than explicit contradiction.

Variation by drawing complexity was most pronounced for E4. The pooled E4 rate increased
monotonically with student representational competence level, from 0.13 errors/feedback for Level~1
drawings to 0.33 for Level~2 and 0.50 for Level~3. In contrast, pooled contradiction-type errors
(E1--E3) were comparatively stable across levels, rising only modestly from 0.17 (Level~1) and 0.19
(Level~2) to 0.24 (Level~3). Thus, as drawings became more complex and information-dense, modal
decoupling increasingly manifested as false absence rather than as explicit mismatches. This pattern
is consistent with the expectation that more information-dense drawings offer more opportunities for
false absence errors: as the number of depicted elements increases, the probability of the model
overlooking at least one of them rises.

Task-level analyses revealed complementary boundary conditions for the two decoupling forms. For E4,
the highest pooled rates occurred in R4 (water condensation; 0.500) and R3 (heating a solid; 0.414),
with similarly elevated rates in R1 (red dye diffusion; 0.396). The lowest pooled E4 rate occurred
in R5 (boiling water; 0.188). For contradiction-type errors (E1--E3), the highest pooled rates
occurred in R2 (balloon expansion/contraction; 0.40) and R1 (0.396), whereas R5 again showed the
lowest contradiction-type rate (0.10). Taken together, these patterns indicate that modal decoupling
is not uniformly distributed across scientific contexts: some tasks preferentially elicit
contradiction-type failures, whereas others elicit false absence, and both error forms are least
frequent in R5.

Although RQ2 focuses on pooled behavior, the same qualitative pattern was observed within each
workflow: E4 rates increased with drawing complexity in both C1 and C2, and the tasks with the
highest pooled contradiction-type error rates (R1, R2) remained comparatively high within each
workflow.

\subsection{Persistence Under Inventory-List-First Workflow (Mitigation Probe) (RQ3)}

RQ3 examined whether modal decoupling persists when feedback generation is preceded by an explicit
visual inventory list (C2). Persistence was assessed at the instance level (whether feedback
contained at least one coded grounding failure, E1--E4) and at the error level (mean coded errors
per instance and error-type composition).

Despite the enforced inventory step, modal decoupling remained substantial. Under C2, 33.3\% of
feedback were flawed (50/150; 95\% CI [26.3\%, 41.2\%]), meaning they contained at least one
grounding failure (E1--E4). For context, the corresponding flawed-instance rate in the direct
workflow was 49.3\% (74/150; 95\% CI [41.4\%, 57.3\%]). Thus, C2 reduced but did not eliminate
pedagogically invalid feedback, with approximately one in three outputs remaining invalid. Compared
to C1, C2 reduced the flawed-instance rate by 16.0 percentage points, but did not eliminate invalid
feedback.

Error rates per instance showed the same pattern. Collapsing across all error categories, mean
errors decreased from 0.654 errors/instance in C1 to 0.414 errors/instance in C2. Reductions
occurred in contradiction-type errors (E1--E3: 0.267 to 0.147 errors/instance) and false absence
errors (E4: 0.387 to 0.267 errors/instance). These findings indicate that substantial error burden
persists under C2 even when the MLLM is prompted to explicitly inventory visual evidence before
generating feedback.

Table~\ref{tab:decoupling} summarizes the number of feedback in which each modal decoupling failure
type was present (binary present/absent coding) by workflow. Because multiple error types can
co-occur in a single feedback, counts are not mutually exclusive.

\begin{table}
\centering
\caption{Frequency of modal decoupling types by workflow.}\label{tab:decoupling}
\begin{tabular}{lllll}
\hline
Error type & C1: $n$ (\%) & C2: $n$ (\%) & $\Delta$ Risk (\textit{pp}) & $p$-Value (McNemar) \\
\hline
E1 (object mismatch) & 7 (4.7) & 8 (5.3) & 0.7 & 1.00 \\
E2 (attribute mismatch) & 11 (7.3) & 4 (2.7) & $-$4.7 & .039\,* \\
E3 (relation mismatch) & 22 (14.7) & 10 (6.7) & $-$8.0 & .017\,* \\
E4 (false absence) & 58 (38.7) & 40 (26.7) & $-$12.0 & .003\,* \\
Any flaw (E1--E4) & 74 (49.3) & 50 (33.3) & $-$16.0 & .001\,* \\
\hline
\end{tabular}\\[4pt]
{\footnotesize\textit{Note.} Values indicate the number of feedback in which the error type was
present (binary coding: absent $= 0$, present $= 1$). A single feedback may contain multiple error
types. $p$-values were computed using McNemar's test on paired per-instance indicators (C1 vs C2; $N
= 150$ drawings). * $p < .05$.}
\end{table}

Under C2, false absence (E4) remained the dominant residual failure mode (40 instances), exceeding
contradiction-type errors (E1--E3 combined: 22 instances). Paired comparisons indicated that C2
reduced the prevalence of E2--E4 (E2: 11 to 4, $p = 0.039$; E3: 22 to 10, $p = 0.017$; E4: 58 to 40,
$p = 0.003$), whereas E1 did not change reliably (7 to 8, $p = 1.000$). Taken together, the
inventory-list-first workflow shifted the composition of grounding failures but did not eliminate
modal decoupling: substantial residual errors remained, including both contradiction-type failures
and false absence relative to the student's drawing.

As a descriptive check, error density per generated word decreased from 0.0083 in C1 to 0.0045 in
C2. This pattern suggests that reductions observed under C2 are not solely an artifact of
differences in response length, although error density per word is not treated as a primary
grounding validity metric.

Overall, the inventory-list-first workflow reduced the flawed-instance rate from 49.3\% to 33.3\%
and lowered mean errors per instance (0.654 to 0.414), but did not suppress modal decoupling.
Residual failures under C2 remained common, and false absence (E4) continued to dominate the
remaining error profile, indicating that explicit inventorying does not reliably bind feedback to
the student's visual evidence.

\subsection{Linguistic Indicators and Modal Decoupling (RQ4)}

RQ4 tested whether modal decoupling is detectable from the feedback text, that is whether feedback
that sounds more visually grounded (more concrete references, more spatial language, fewer hedges)
is less likely to contain grounding failures.

Across both workflows, non-flawed and flawed feedback looked very similar on the measured text
features. In C2, flawed feedback was only slightly longer than non-flawed feedback (91.4 vs.\ 88.6
words). Lexical densities differed only modestly: physical-term density was 0.44\% for flawed vs.\
0.52\% for non-flawed feedback, spatial-term density was 0.20\% vs.\ 0.26\%, and hedging was 0.24\%
vs.\ 0.20\%. In C1, flawed feedback was also slightly longer (78.1 vs.\ 74.6 words) and showed lower
spatial-term density (0.16\% vs.\ 0.32\%) and hedging (0.06\% vs.\ 0.11\%) than non-flawed feedback.
Overall, more visually grounded language was not associated with fewer errors.

Cluster-robust logistic regression models, controlling for modeling task, drawing level, and
workflow condition, showed no reliable associations between the presence of grounding failures and
physical-term density, spatial-term density, or hedging density (all $p > .34$).
The only consistent text-related signal was length: higher word count was associated with higher
odds of false absence errors (E4) ($OR = 1.046$ per word, $p = .017$), while length was not
associated with contradiction errors (E1--E3) ($p = .601$).

Adding linguistic style features to a baseline model (task, level, condition) improved
discrimination only negligibly (AUC 0.746 vs.\ 0.755; $\Delta$AUC $= 0.009$). Consistent with this,
a ``grounding style'' quartile split within C2 showed no reliable trend in flaw rates (linear trend
$p = 0.572$), and quartile contrasts were non-significant (all $p \geq 0.215$). Surface linguistic
cues provide little to no practically useful signal for identifying flawed feedback.

\section{Discussion}

This study provides empirical evidence of grounding invalidity in an applied setting: under two
deployment-realistic workflows, off-the-shelf MLLM feedback on student scientific drawings is
frequently not grounded in the visual evidence. Inventory-list-first reduces the overall error rate
relative to direct feedback generation, but it does not make feedback valid. The dominant failure
mode is false absence: feedback often states that depicted elements are missing or redundantly
requests their addition. As a result, feedback can look pedagogically plausible in form while being
decoupled from the student's drawing.

\subsection{Modal Decoupling as an Outcome of Known Technical Constraints (RQ1 \& RQ2)}

Our theoretical framing treats modal decoupling as a plausible functional outcome of known
constraints in current MLLMs, including geometric separation between modalities, optimization bias
toward language, and visual grounding decay. The prevalence and structure of grounding failures
observed here are consistent with that framing. First, modal decoupling was common under both
workflows, indicating that grounding failures are not limited to a single prompting style. Second,
the error profile was dominated by false absence (E4), which increased with drawing competence
level. This pattern suggests that the main bottleneck is not only identifying visual elements, but
maintaining coverage and evidence selection when drawings become information-dense and relationally
complex. Task-level variation further supports that decoupling is shaped by representational
demands: some contexts preferentially caused mainly contradiction-type errors (E1--E3), while others
showed especially high false absence (E4). Together, these boundary conditions align with the view
that MLLMs fall back to high-probability pedagogical patterns when visual signal is weak, vague, or
difficult to sustain throughout generation.

\subsection{Why Inventory-List-First Helps, and Why It Plateaus (RQ3)}

RQ3 tested whether an inventory-list-first workflow meaningfully mitigates grounding failures caused
by modal decoupling. The paired comparisons indicate that it reduces several error categories,
especially attribute mismatch (E2), relation mismatch (E3), and false absence (E4). This is
compatible with the intended function of the inventory step: to frontload observation and keep
evidence accessible during evaluation. However, grounding failures persisted at a substantial level
under inventory-list-first, with roughly one in three outputs still flawed. The remaining error
profile under C2 was still dominated by E4, and object mismatch (E1) did not change reliably. This
combination suggests a ceiling for what inventory prompting can accomplish: it can improve some
downstream specificity once entities are ``locked in,'' but it does not correct early failures in
establishing a correct object set. In short, inventory-list-first behaves as a risk-reduction
heuristic, not as a grounding validity mechanism.

\subsection{``Sounding Grounded'' Is Not the Same as Being Grounded (RQ4)}

RQ4 asked whether grounding validity is detectable from feedback text properties that are often
taken as proxies for visual grounding. As motivated in Section~2.3, if the surface linguistic
profile of feedback cannot distinguish grounded from ungrounded outputs, then basic filters on
text-based characteristics fail to catch the errors this study documents. In our data, linguistic
surface cues provided little diagnostic value. Flawed and non-flawed feedback were similar in length
and in grounding-oriented lexical densities, and adding these features only negligibly improved
discrimination beyond task, competence level, and workflow condition. This result challenges a
common implicit safety assumption: that more concrete references, more spatial language, or more
hedging indicates better alignment with the drawing. Here, feedback can read as specific and
well-formed while still containing grounding failures, implying that fabricated utility is difficult
to detect from the feedback text alone. For practice, this means that systems cannot rely on
stylistic signals as a substitute for explicit grounding checks.

\subsection{Educational and Practical Implications for Building and Using MLLM Feedback Systems}

Given our results, prompting should be framed as risk reduction, not validation. If feedback
validity entails grounding in the drawing, then systems require mechanisms that make evidence a
binding constraint rather than an optional preface. This points away from ``generate then justify''
\cite{ref_ji} and toward ``generate only what can be grounded,'' with explicit support for
verification and abstention when grounding is uncertain. The dominance of E4 (false absence) is
important: feedback can look responsive while claiming that depicted elements are missing or
redundantly requesting elements that are already present, especially in information-dense drawings.
Valid feedback systems will likely require structured intermediate representations of depicted
entities, attributes, and relations (e.g., semantic graph-style representations) \cite{ref_latif},
along with explicit checks that each feedback claim maps back to that representation and its
supporting evidence, because neither inventory-list-first strategies nor linguistic signals offer a
dependable correctness screen.

When feedback is driven by model priors instead of the student's drawing, it introduces
construct-irrelevant variance into feedback on students' representational competence. The increase
of E4 (ignoring evidence) with competence level adds a specific concern: the system may fail most
where student work is richest and most diagnostic, potentially disadvantaging more advanced drawings
by ignoring relevant structure and requesting redundant additions. When feedback treats
already-depicted elements as absent, it may prompt students to revise work that correctly represents
the target phenomenon, undermining the instructional function of formative feedback.

\section{Conclusion}

The results demonstrate that off-the-shelf MLLMs frequently generate feedback on student drawings that seems
pedagogically plausible yet is visually unanchored, confirming the validity risk of grounding failures
potentially due to modal decoupling where linguistic priors override specific visual evidence.
Inventory-list-first improves reduces some grounding failures without completely resolving them.
The path to valid feedback therefore might not lie not primarily in more elaborate prompting, but in
workflows and system designs that treat visual evidence as a binding constraint on what feedback may
claim. Consequently, prompt-level scaffolds with off-the-shelf MLLMs are insufficient for valid
autonomous feedback; such systems should only operate under human oversight rather than function as
direct feedback agents for students.

\begin{credits}
\subsubsection{\ackname}
The research reported here was supported by the Institute of Education Sciences, U.S. Department of
Education, through Grant R305C240010 (PI Zhai). The opinions expressed are those of the authors and
do not represent views of the Institute or the U.S. Department of Education.

\subsubsection{\discintname}
The authors have no competing interests to declare that are relevant to the content of this article.
\end{credits}


\end{document}